\pdfoutput=1

\documentclass[11pt]{article}

\usepackage{natbib,upgreek}
\usepackage{amsmath}
\usepackage{graphicx,psfrag,epsf}
\usepackage{enumerate}
\usepackage{natbib}
\usepackage{subfig}
\usepackage{amsmath}
\usepackage{amsmath, amssymb, amsfonts}
\usepackage{mathtools}
\usepackage{nameref}
\usepackage{sectsty}
\usepackage{lipsum}
\mathtoolsset{showonlyrefs=true}
\usepackage[a4paper,top=2cm,bottom=2cm,left=2cm,right=2cm]{geometry}

\begin{document}

	\title{\bf Bayesian Hidden Markov Modelling Using Circular-Linear General Projected Normal Distribution}
	\author{Gianluca Mastrantonio
		\thanks{gianluca.mastrantonio@uniroma3.it Corresponding author}
		\hspace{.2cm}\\
		Department of Economics, University of Romatre\\
		and \\
		Antonello Maruotti \thanks{a.maruotti@soton.ac.uk } \\
		Southampton Statistical Science Research Institute, University of Southampton\\
		and \\
		Giovanna Jona Lasinio \thanks{giovanna.jonalasinio@uniroma1.it}\\
		Department of Statistical Science, Sapienza University of Rome
	}
	
%
%
\date{}
\maketitle
	\begin{abstract}
		We introduce a multivariate hidden Markov model to jointly cluster time-series observations with different support, i.e. circular and linear. Relying on the general projected normal distribution, our approach allows for bimodal and/or skewed cluster-specific distributions for the circular variable. Furthermore, we relax the independence assumption between the circular and linear components observed at the same time. Such an assumption is generally used to alleviate the computational burden involved in the parameter estimation step, but it is hard to justify in empirical applications. We carry out a simulation study using different data-generation schemes to investigate model behavior, focusing on well recovering the hidden structure. Finally, the model is used to fit a real data example on a bivariate time series of wind speed and direction.
	\end{abstract}
	\maketitle

	\section{INTRODUCTION}
	\label{s:intro}
	
	Hidden Markov models (HMMs) have become more frequently used to provide a natural and flexible framework for univariate and multivariate time-dependent data (e.g. time-series, longitudinal data). They are a class of mixture models in which the data-generation distribution depends on the state of an underlying and unobserved Markov process. Hidden Markov modelling has been used as a statistical tool for density estimation \citep[][]{Langrock2014,Dannemann2012}, supervised and unsupervised classification \citep[][]{lagona2012,   alfo2010,fruhwirthschnatter2006}    and a wide range of empirical problems in environmetrics  \citep[][]{Martinez2013,Langrock2012}, medicine  \cite[][]{Langrock2013,Lagona2014}, education  \cite[][]{bartolucci2011}.   For a comprehensive introduction to fundamental theory of HMMs encountered in practice, see the review papers of \cite{bartolucci2014}, \cite{maruotti2011} and monographs by \cite{bartolucci2012}, \cite{zucchini2009b} and \cite{cappe2005}.
	
	The literature on multivariate hidden Markov modelling is dominated by Gaussian HMMs \citep[][]{spezia2010,Bartolucci2010c, Geweke2011}. Modelling multivariate time series with non-normal components of mixed-type is challenging. The joint distribution of multivariate (mixed-type) data is usually specified as a
	mixture having products of univariate distributions as components \citep[see e.g][]{lagonaet2011,lagona2011b,Zhang2010}. \citet{bartolucci2009} is a notable exception. In other words, random variables are assumed conditionally independent given the latent structure. Although conditional independence facilitates parameters estimation, it is a too restrictive assumption in many empirical applications and may not properly accommodate for the complex shape of multivariate distributions \citep{Baudry2010}. Moreover, an unnecessary number of latent states is often needed to obtain reasonable  fit, at the price of an increased computational burden and difficulties in interpreting results, as shown in the simulation study section.
	
	In this paper, we propose a bivariate distribution for circular-linear time-series in a HMM framework. We accommodate for nonstandard features of data including correlation in time and across variables, mixed supports (circular and linear) of the data, the special nature of circular measurements, and the occurrence of missing values. 
	We relax the conditional independence assumption between circular and linear variables by taking a fully parametric approach. 
	
	This is not the first attempt to jointly modelling circular and linear variables in a HMM framework. 
	\cite{Bulla2012} introduced a latent-class approach to the analysis of multivariate mixed-type data by assuming that circular and linear variables are conditionally independent given the states visited by a latent Markov chain while \cite{kato2008} propose a hyper-cylindrical distribution.  The latter is problematic (in the HMM setting in particular), because little is known about efficient
	estimation procedures and identifiability issues under hyper-cylindrical parametric
	models. In addition, mixtures of hyper-toroidal densities would group data according
	to clusters of difficult interpretation, without necessarily improving the fit of the model.
	
	We introduce a flexible structure, relying on the general projected normal distribution \cite[][]{wang2014}, to model circular measurements, and extending \cite{Bulla2012} to a more general setting, allowing for (conditional) correlation between circular and linear variables. We treat the circular response as projection onto the unit circle of a bivariate variable and define the joint circular-linear distribution through the specification of a  multivariate model in a multivariate linear setting,
	extending \cite{wang:2014} to a clustering framework. 
	
	The resulting hidden Markov model parameters are estimated in a Bayesian framework. 
	We provide details on how to  fit the model  by using  MCMC methods,  and we point out possible drawbacks in the implementation of the algorithm. Advantages of the Bayesian approach, with respect to the EM algorithm, include a convenient framework to simultaneously account  for several data features, adjust for identifiability issues and produce natural measures of  uncertainty for model parameters.
	For a general discussion see e.g. \cite[][]{ryden1998,ryden2008,yilidirim2014}.

	We illustrate the proposal  by a large-scale simulation study in order to investigate the empirical behaviour of
	the proposed approach with respect to several factors, such as the number of observed times,
	{the association structure between the circular and linear variables} and the fuzziness of the classification. We evaluate model performance in recovering the true model structure, we compare several models on the
	basis of their 
	ability to accurately estimate the vectors of state-dependent parameters and hidden parameters.
	Finally, we test the proposal by analysing time series of semi-hourly wind directions and speeds, recorded in the period 12/12/2009, 12/1/2010 by the buoy of Ancona, located in the Adriatic Sea at about 30 km from the coast. 
	
	The rest of the paper is organized as follows. In Section \ref{s:prel},
	we briefly review relevant aspects necessary for the introduction of our approach and outline some results about the projected normal distribution. Section
	\ref{s:clhmm} discusses the specification of the circular-linear general projected normal hidden Markov model and  provides Bayesian inference. Computational details and parameters estimation are discussed as well.
	Section \ref{s:ss} presents a large-scale simulation study. In Section \ref{s:real}, the application
	of the proposed methodology is illustrated through a
	real-world data set. Some concluding remarks are given in
	Section \ref{s:discussion}.
	
	\section{PRELIMINARIES} \label{s:prel}

	Circular data are a particular class of directional data, specifically, they are
	directions in two dimensions. {To} analyze circular data is challenging because usual statistics,
	which have been developed for  linear data (for example  the mean and variance), will not be meaningful and will
	be misleading when applied to  directional data without taking into account the particular definition of the domain. 
	There are many ways to define distributions in a circular domain, see the book of \cite{Merdia1999} for a comprehensive overview. The one we used in this paper   is to radially
	project onto the circle a probability distribution originally defined on the plane. Let ${\bf Z} = [Z_1, Z_2]^{\prime}$ be a 2-dimensional random vector such that $\Pr({\bf Z}={\bf 0})=0$. Then, its radial projection 
	${\bf W}= \left[ \begin{array}{c} W_1 \\ W_2    \end{array}\right] = \frac{{\bf Z}}{||{\bf Z}||}$
	is a random vector on the unit circle, which can be converted to a  random angle $X$ relative to some direction treated as 0 via the transformation $X = \arctan^*  \frac{W_2}{W_1} =  \arctan^*  \frac{Z_2}{Z_1} \in [0, 2 \pi) $, where  the function $\arctan^*$ is a quadrant specific inverse of the tangent function, sometimes called \emph{atan2}, that takes into account the signs of $W_1$ and $W_2$ to identify the right quadrant of $X$; for a formal definition see \cite{Jammalamadaka2001}, pag. 13. Note that ${\bf W}= \left[ \begin{array}{c} \cos X \\ \sin X    \end{array}\right]$ and let $R = ||Z ||$ the following relation holds:
	$\left[\begin{array}{cc} Z_{1}\\ Z_{2} \end{array}\right] = \left[\begin{array}{cc} R\cos X\\ R \sin X \end{array}\right] = R \mathbf{W} $.  
	
	By assuming  ${\bf Z}\sim N_2(\cdot|\boldsymbol{\mu},\boldsymbol{\Sigma})$,  with $\boldsymbol{\Sigma}=\left[\begin{array}{cc}\sigma^2_1&\sigma_1\sigma_2\rho\\ \sigma_1\sigma_2\rho & \sigma^2_2 \end{array}\right]$ and $\boldsymbol{\mu} = \left[\begin{array}{c} \mu_1\\ \mu_2 \end{array}\right] $,  $X$ is said to have a 2-dimensional projected normal distribution, denoted by $PN_2(\cdot| \boldsymbol{\mu},\boldsymbol{\Sigma})$. Since the distribution of $X$ does not change if we multiply $\mathbf{Z}$ for a positive constant $c>0$,  for 
	identifiability purposes, following \cite{Wang2013},  $\sigma_2^2$ is set to be 1. The projected normal distribution is  specified  as a four parameters distribution:  $PN_2(\cdot| \mu_1, \mu_2, \sigma_1^2, \rho)$. 

	\begin{figure}[t!]
		\centering
		\includegraphics[scale=0.5]{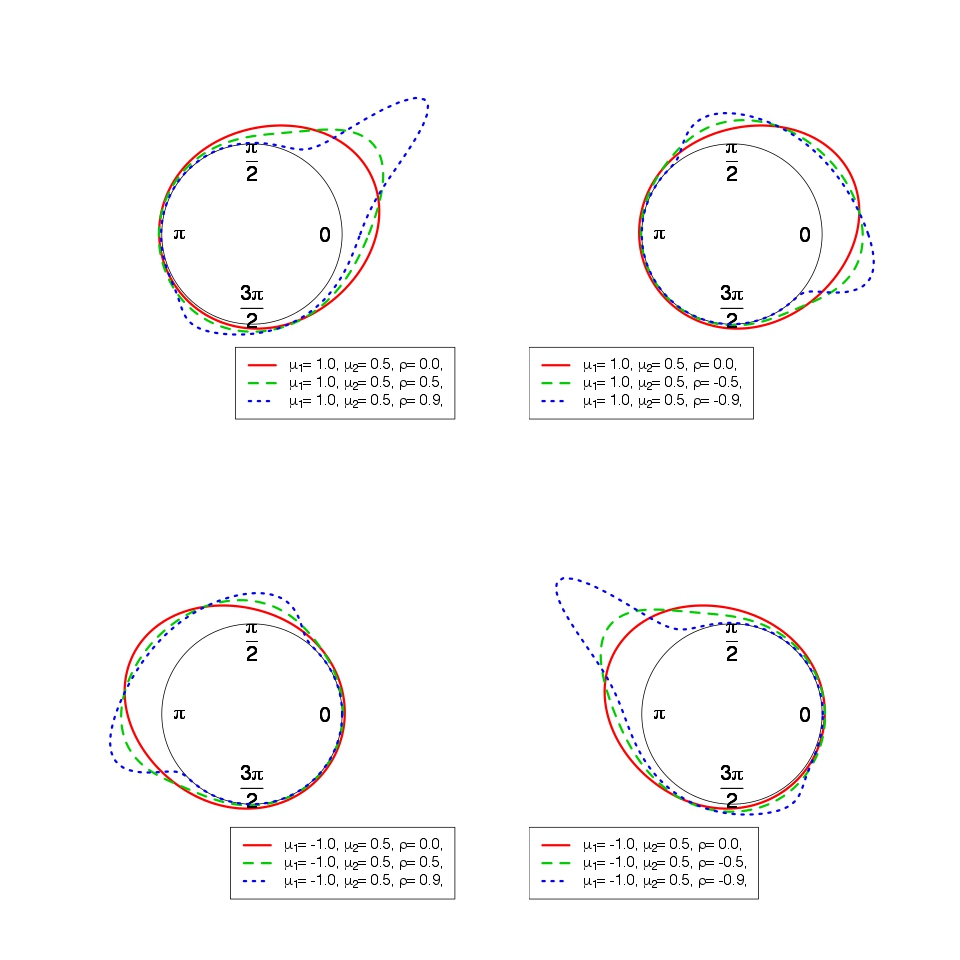}  
		\caption{Shape of the projected normal distribution for $\sigma^2=1$ and different values of $\rho$, $\mu_1$ and $\mu_2$} \label{fig:rho}
	\end{figure}
	
	We provide some examples to illustrate the flexibility of the  projected normal (PN) distribution.
	The PN density can be symmetric,  asymmetric or
	possibly bimodal and apart form some special case, the interpretation of the parameters can be difficult. The number of modes and the shape depend on the value of the all parameters and different sets of parameters can give really similar shapes. 
	As a general comment we highlight that $\mu_1$ and $\mu_2$  are the means of the two Cartesian coordinates $z_1$ and $z_2$ and are respectively connected to the cosine and sine of the circular variable.   By fixing $\sigma_1^2=1$ and $\rho=0$,  the resulting distribution is unimodal and symmetric and if $\mu_1=\mu_2=0$  the distribution becomes a circular uniform. Departure from  zero for the two means, in the case of  identity covariance matrix,  creates one  mode in the trigonometric  quadrant with the same sign of the means, e.g. if $\mu_1>0$ and $\mu_2<0$ then the mode is in the quadrant with positive cosine and negative sine; higher values of a mean attract the mode to its correspondent  axis. 
	By allowing the $\rho$ parameter to vary, we obtain very flexible shapes. The resulting distribution shows asymmetry with more mass of probability near the axis with the highest $\mu$.  By increasing $|\rho|$, bimodality is detected, Figure \ref{fig:rho}. Moreover, for $\sigma_1^2<1$ the modes are closer to the sine axis; while for $\sigma_1^2>1$ the modes are closer to the cosine axis.

	\section{THE CIRCULAR-LINEAR GENERAL PROJECTED NORMAL HIDDEN MARKOV MODEL}
	\label{s:clhmm}
	
	\subsection{The Model}
	Let $\mathcal{T} = \{0,1,\dots, T-1, T\}$, in this paper we consider a bivariate  time series $[\mathbf{x}, \mathbf{y}] = \{ [x_t, y_t]; t \in \mathcal{T} \backslash \{ 0  \} \}$ with circular, $x_t$, and linear, $y_t$, components.  Our aim is to jointly classify $[x_t, y_t]$ in $K$ classes, generally called regimes or states, with a HMM-based classifier.\\
	Let $\pi_{k,h}$ indicates the probability to move from state $k$ to state $h$ and let $\xi_{tk}$ be an indicator variable such that if we are in state $k$ on time $t$ it is 1, otherwise is 0. Then $f(\xi_{th}=1| \xi_{t-1k}=1)= \pi_{k,h}$ and we set $f(\xi_{0k}) = \pi_{k}$.  
	We indicate with 
	\begin{equation}
	\boldsymbol{\pi}=
	\left[ \begin{array}{llll}
	\pi_{1,1} & \pi_{1,2} & \cdots &  \pi_{1,K}\\
	\pi_{2,1} & \pi_{2,2} & \cdots &  \pi_{2,K}\\
	\cdots  & \cdots & \cdots &  \cdots  \\
	\pi_{K,1} & \pi_{K,2} & \cdots &  \pi_{K,K}
	\end{array} \right], \, \sum_{h=1}^K \pi_{k,h}=1, \, k=1,2, \dots , K,
	\end{equation}
	the transition matrix that governs the evolution of the Markov chain, $\boldsymbol{\pi}_0 = \left[ \pi_{1}, \pi_{2}, \dots , \pi_{K}\right]^{\prime}$  and $\boldsymbol{\xi} = [\boldsymbol{\xi}_0, \boldsymbol{\xi}_1, \dots , \boldsymbol{\xi}_T]^{\prime}$ where $\boldsymbol{\xi}_t = [\xi_{t1}, \xi_{t2}, \dots , \xi_{tK}]^{\prime}$. 
	
	Let $n_{k,h}= \sum_{t=1}^T \xi_{t-1k }\xi_{th}$ be the number of times we move from state $k$ to  state $h$,  the joint density of the vector of states is $f(\boldsymbol{\xi} |  \boldsymbol{\pi}, \boldsymbol{\pi}_0) =\prod_{k=1}^K  \pi_{k}^{\xi_{0k}}\prod_{k=1}^K \prod_{h=1}^K \pi_{k,h}^{n_{k,h}}$.
	In the literature on HMM for circular-linear variables, see for example \cite{Bulla2012} and \cite{Holzmann2006}, it is generally assumed that conditioning to the latent vector $\boldsymbol{\xi}$,  the pairs $[x_t, y_t] $ and $[x_g,y_g]$ are independent if $g \neq t$ and at the same time $x_t \perp y_t$. As a result, the conditional distribution of the observed process, given the latent process, takes the form of a product density, say
	$f(\mathbf{x}, \mathbf{y}| \boldsymbol{\xi}) = \prod_{k=1}^K\prod_{t=1}^T \left[ f({x}_t| \xi_{tk}=1) f({y}_t| \xi_{tk}=1)\right]^{\xi_{tk}}.$
	We maintain the so-called conditional independence property: given the hidden state at time $t$, the distribution of the observation at this time is fully determined, but we relax the assumption on independence between the circular and linear variables observed at the same time. Thus, we get a multivariate conditional distribution
	$f(\mathbf{x}, \mathbf{y}| \boldsymbol{\xi}) = \prod_{k=1}^K\prod_{t=1}^T   f({x}_t, y_t| \xi_{tk}=1)^{\xi_{tk}}.$ \\
	
	Let
	$\mathbf{Z}_t| \xi_{tk}=1\sim N \left( \boldsymbol{\mu}_k    ,   \boldsymbol{\Sigma}_k    \right)$, with  $\mathbf{Z}_t = \left[\begin{array}{cc} Z_{t1}\\ Z_{t2} \end{array}\right] $,  $\boldsymbol{\mu}_k=\left[\begin{array}{cc} \mu_{k1}\\ \mu_{k2} \end{array}\right]$ $\boldsymbol{\Sigma}_k=\left[\begin{array}{cc} \sigma_{k1}^2 & \sigma_{k1} \rho_k \\  \sigma_{k1} \rho_k & 1 \end{array}\right]$ and let $R_t = || \mathbf{Z}_t ||$. We define $X_t$ as the radial projection of $\mathbf{Z}_t$: $X_t =   \arctan^*  \frac{Z_1}{Z_2}$ and then $X_t| \xi_{tk}=1\sim PN_2 (\boldsymbol{\mu}_k, \boldsymbol{\Sigma}_k)$. We can write easily the joint density of $[X_t,R_t]$ that is the density that arises by a  variable transformation from the bivariate normal $Z_t$ to its polar system representation. Let $\phi_h(\zeta | \mathbf{M},\mathbf{V} )$  be the probability density function of a h-variate normal distribution with mean $\mathbf{M}$ and covariance matrix $\mathbf{V}$ evaluated in $\zeta$, then  
	\begin{equation} \label{eq:rx}
	f({x}_t, r_t| \xi_{tk}=1) = \phi_2(r_t \mathbf{w}_t | \boldsymbol{\mu}_k,\boldsymbol{\Sigma}_k ) r_t
	\end{equation}

	We built the (conditional) joint density $f({x}_t, y_t| \xi_{tk}=1)= f( y_t|{x}_t, \xi_{tk}=1)f({x}_t| \xi_{tk}=1)$  as a marginalization over the latent variable $R_t$:  
	$f({x}_t, y_t| \xi_{tk}=1) = \int_{r_t}  f( y_t|{x}_t, r_t,\xi_{tk}=1) f({x}_t, r_t| \xi_{tk}=1) d r_t$,
	where $f({x}_t, r_t| \xi_{tk}=1)$ is specified in equation \eqref{eq:rx}    and $y_t | x_t, r_t, \xi_{tk}=1$ is defined through a circular-linear regression.  However, there is not an obvious and standard  way to formalize the relations between circular and linear variables. \cite{Jammalamadaka2001} propose a flexible approach using  trigonometric polynomials  while \cite{mardia1976} and  \cite{Johnson1978} proposed a regression where the covariates are the sine and cosine components of the circular variable. Here, following  \cite{wang:2014}, we specify the relation as $y_t= \gamma_{k0}+\gamma_{k1}r_t \cos x_t+\gamma_{k2}r_t\sin x_t+\epsilon_{tk}$, with $\epsilon_{tk} \sim N(0,\sigma_{ky})$. Thus,    
	$y_t | x_t, r_t, \xi_{tk}=1$ is distributed as a normal variable with mean   $ \gamma_{k0}+\gamma_{k1}r_t \cos x_t+\gamma_{k2}r_t \sin x_t$ and variance $\sigma_{ky}^2$. Note that the regression can be seen as a linear regression between $y_t$ and the inline variables $r_t \cos x_t$ and $r_t \sin x_t$.  This type of representation gives more flexibility to the circular linear regression than the ones proposed by \cite{mardia1976} and  \cite{Johnson1978}. Notice that with the $r_t$ variable,  for a given value of the circular variable at different time point, say $x_t = x_{t^{\prime}}, t \neq t^{\prime}$, the relation between $x_t$ and $y_t$ and $x_{t^{\prime}}$ and $y_{t^{\prime}}$ can be different as it depends on the realization of the non observed variable $r_t$.

	Then we have  that 
	\begin{equation}  \label{eq:jointcl}
	f(x_t,y_t |\xi_{tk}=1) =\frac{ \phi_1(y_t|\gamma_{k0},\sigma_{ky}^2  ) \phi_2(\boldsymbol{\mu}_k|\mathbf{0}_2, \boldsymbol{\Sigma}_{k})  \left[   m_{tk} \Phi\left(\frac{m_{tk}}{\sqrt{v_{tk}}}\right) + \phi_1\left(  m_{tk} |0,v_{tk} \right)     \right]}{\phi_1\left({m_{tk}}|0, v_{tk}   \right)} , 
	\end{equation}
	where   $\Phi$ is the cumulative density function of a standard normal distribution, $ \mathbf{w}= \left[\begin{array}{c}
	\cos x_t  \\ \sin x_t
	\end{array}\right]$,   $v_{tk}= \left[ \frac{  c_{tk}^2}{\sigma_{ky}^2}+ \mathbf{w}_t^{\prime}\Sigma_{k}^{-1}\mathbf{w}_t     \right]^{-1}$,  $m_{tk}=v_{tk} \left[\frac{c_{tk}(y_t-\gamma_{k0})}{\sigma_{ky}^2 } +  \mathbf{w}_t^{\prime}\Sigma_{k}^{-1}\boldsymbol{\mu}_k \right]$ and $c_{tk} =  \mathbf{w}_t^{\prime} \left[\begin{array}{c}
	\gamma_{k1}  \\ \gamma_{k2}
	\end{array}\right]$.  
	The  Circular Linear General projected normal (CL-GPN) distribution with parameters $[{\mu}_{k1},{\mu}_{k2},{\sigma}_{k1}^2,{\rho}_k, {\gamma}_{k0},{\gamma}_{k1},$  ${\gamma}_{k2}, {\sigma}_{ky}^2]^{\prime}$ is thus defined in \eqref{eq:jointcl}.
	In this setting the parameter $\gamma_{k1}$ and $\gamma_{k2}$ govern the dependency between the two variables (linear and circular), $\gamma_{k1}$ is connected to the  correlation between the linear variable and the cosine of the circular, $\gamma_{k1}$ is connected to the correlation between the linear variable and the sine of the circular.

	\cite{wang:2014} and \cite{Wang2013} argue that working with the projected normal density is not easy and its form is practically intractable (to see  the closed form of the PN density see \cite{Wang2013}). Since the CL-GPN is based on the PN, is itself an intractable distribution
	and  the implementation of the MCMC algorithm can be difficult.
	However  the introduction of  $r_t$  is of practical use as it simplifies the implementation of the MCMC algorithm, see Section \ref{sec:comp}. 

	%
	\ 
	\subsection{Posterior Inference} \label{sec:postinf}
	Let $\boldsymbol{\Psi}$ be the vector of all the parameters of the CL-GPN in all the $K$ regimes, we have the following posterior distribution
	\begin{equation} \label{eq:post4}
	f\left( \boldsymbol{\pi}, \boldsymbol{\xi},\boldsymbol{\pi}_0, \boldsymbol{\Psi}, \mathbf{r}| \mathbf{x}, \mathbf{y}  \right) =  \frac{f \left(\mathbf{r}, \mathbf{x} , \mathbf{y}| \boldsymbol{\Psi}, \boldsymbol{\xi}   \right)   f\left( \boldsymbol{\xi}_{-0}| \boldsymbol{\xi}_{0}, \boldsymbol{\pi} \right)    f\left(  \boldsymbol{\pi}\right)  f\left( \boldsymbol{\xi}_{0}| \boldsymbol{\pi}_0 \right)       f\left(\boldsymbol{\pi}_0  \right)   f\left( \boldsymbol{\Psi}\right)    }{f\left(  \mathbf{x}, \mathbf{y}\right)}
	\end{equation}
	where $\mathbf{r}= [r_1,\dots , r_t]^{\prime}$ and $f(\mathbf{x}, \mathbf{y}, \mathbf{r}| \boldsymbol{\xi}) = \prod_{k=1}^K\prod_{t=1}^T f({x}_t, r_t, y_t| \xi_{tk}=1)^{\xi_{tk}}$.
	As prior distribution we assume: 
	${\mu}_{ki} \sim N(\cdot, \cdot)$, $ \sigma_{k1}^2  \sim IG(\cdot, \cdot)$, $ \rho_{k}  \sim N(\cdot, \cdot)I(-1,1)$, $\sigma_{ky}^2  \sim IG(\cdot, \cdot)$, $  \gamma_{kj}\sim N(\cdot, \cdot)$ for $k=1,\dots , K, \, i = 1,2, \,j=1,2,3$, where  $IG(\cdot,\cdot)$  indicates the Inverse Gamma distribution,  $\boldsymbol{\pi}_{0} \sim Dir(\cdot)$ and $\boldsymbol{\pi}_{k,.} \sim Dir(\cdot)$ where  $Dir(\cdot)$ indicates the Dirichlet distribution and $\boldsymbol{\pi}_{k,.}$ is the $k^{th}$ row of $\boldsymbol{\pi}$: we assume $\boldsymbol{\pi}_k \perp \boldsymbol{\pi}_{k^{\prime}}$ if  $k \neq k^{\prime}$.
	The prior specification allows us to marginalized over $\boldsymbol{\pi}$ and $\boldsymbol{\pi}_0$ reducing of $K^2$ the number of parameters to simulate and  leads to a more efficient   and stable algorithm (see \cite{Banerjee2003} and Section  \ref{sec:comp}). Note that we can always sample from  $f(\boldsymbol{\pi}_{k,.}| \mathbf{x}, \mathbf{y})=       
	\sum_{\boldsymbol{\xi} }     f(\boldsymbol{\pi}_{k,.}| \boldsymbol{\xi})    f(\boldsymbol{\xi} | \mathbf{x}, \mathbf{y})$
	and
	$f(\boldsymbol{\pi}_{0}| \mathbf{x}, \mathbf{y})= 
	\sum_{\boldsymbol{\xi} }     f(\boldsymbol{\pi}_{0}| \boldsymbol{\xi})    f(\boldsymbol{\xi} | \mathbf{x}, \mathbf{y})$ given the set of $B$ posterior samples  $\{\boldsymbol{\xi}^b\}_{b=1}^B$ of $\boldsymbol{\xi}$,
	with an MCMC integration.   For each sample $\boldsymbol{\xi}^b$ we draw a sample from  $  \boldsymbol{\pi}_{k,.}^b| \boldsymbol{\xi}^b  \sim Dir( \beta+$  $ \sum_{t=2}^T $ $\xi_{t-1k}^b \xi_{t1}^b, $  $ \dots, \beta+ \sum_{t=2}^T \xi_{t-1k}^b \xi_{tK}^b)$ and one from $  \boldsymbol{\pi}_{0}^b | \boldsymbol{\xi}^b  \sim Dir\left( \beta+  \xi_{01}^b, \dots, \beta+ \xi_{0K}^b\right)$. The sets  $\{  \boldsymbol{\pi}_{k,.}^b \}_{b=1}^B$  and $\{  \boldsymbol{\pi}_{0}^b \}_{b=1}^B$ are draw from their respectively marginal posterior distributions.
	The posterior distribution we will work with is then
	\begin{equation} \label{eq:mod1}
	f\left(   \boldsymbol{\xi} , \boldsymbol{\Psi}, \mathbf{r}| \mathbf{x}, \mathbf{y}  \right) =  \frac{f \left(\mathbf{r},  \mathbf{x}, \mathbf{y}| \boldsymbol{\Psi}, \boldsymbol{\xi}   \right)     f\left( \boldsymbol{\Psi}\right)    f(\boldsymbol{\xi}_{-0} | \boldsymbol{\xi}_{0})          f(\boldsymbol{\xi}_0)   }{f\left(  \mathbf{x}, \mathbf{y}\right)}.
	\end{equation}
	where $ f(\boldsymbol{\xi}_{-0} | \boldsymbol{\xi}_{0}) = \int_{\boldsymbol{\pi}}    f\left( \boldsymbol{\xi}_{-0}| \boldsymbol{\xi}_{0}, \boldsymbol{\pi} \right)    f\left(  \boldsymbol{\pi}\right) d \boldsymbol{\pi}$ and  $f(\boldsymbol{\xi}_0) =  \int_{\boldsymbol{\pi}_0}     f\left( \boldsymbol{\xi}_{0}| \boldsymbol{\pi}_0 \right)       f\left(\boldsymbol{\pi}_0  \right)    d \boldsymbol{\pi}_0$ can be computed  in closed form: 
	$f(\boldsymbol{\xi}_{-0} | \boldsymbol{\xi}_{0})= \frac{\Gamma(K\beta)^{K}}{\Gamma\left( \beta\right)^{K^2}}    \frac{ \prod_{k=1}^K\prod_{h=1}^K \Gamma \left(   n_{k,h}+\beta  \right) }{  \prod_{k=1}^K \Gamma \left( n_{k}- \xi_{T,k}+K\beta \right)}   $     
	and
	$f\left( \boldsymbol{\xi}_0 \right) =  \frac{1}{K} $              

	\subsection{Computational Details} \label{sec:comp}
	Model parameters  are estimated with a MCMC algorithm. More precisely the ${\mu}$s, ${\gamma}$s, $\sigma_y^2$s and ${\xi}$ are simulated with a Gibbs sampler while the remaining parameters require the introduction of    a Metropolis step. The full conditionals of $\mu$s and $\gamma$s are normal distributions, those of $\sigma_y^2$s are inverse gamma. The full conditionals for the latent variables $\boldsymbol{\xi}_t, \, t \in \mathcal{T}$ are multinomial and  the vector of probabilities depends on the entire vector of $\boldsymbol{\xi}_{-t} = \boldsymbol{\xi} \backslash  \{  \boldsymbol{\xi}_t\}$.  More precisely, let $s^{-}$ and $s^{+}$ be  the regimes  on time $t-1$ and $t+1$, i.e $\xi_{t-1 s^{-}}=1$ and $\xi_{t+1 s^{+}}=1$ respectively,  let $
	n_{k}^{-t^{\prime}} =\sum_{t=0 \atop t \neq t^{\prime}}^{T} \xi_{tk}
	$ and
	$
	n_{k,h}^{-t^{\prime}} =\sum_{t=1 \atop t \neq t^{\prime},  t \neq t^{\prime}+1}^T \xi_{t-1k }\xi_{t,h}
	$. If $t \in \mathcal{T} \backslash \{0, T \}$.
	\begin{equation}
	f(\boldsymbol{\xi}_t |\mathbf{r}, \mathbf{x}, \mathbf{y}, \boldsymbol{\xi}_{-t},\boldsymbol{\Psi}) \propto \prod_{k=1}^K    \frac{\left(  n_{s^-,k}^{-{ t}}+\beta +a_{s^-,k,s^+} \right)\left(  n_{k,s^+}^{-{ t}}+\beta  \right)    }{ \left(   n_{k}^{-{ t}}- \xi_{T,k}+K\beta \right) }  f( r_{ t},{x}_t, {y_t}| \boldsymbol{\xi}_{t},\boldsymbol{\Psi})         
	\end{equation}
	where $a_{s^-,k,s^+}$ assumes value 1 if $s^-=k=s^+$, 0 otherwise, while
	\begin{equation}
	f(\boldsymbol{\xi}_{ T} \vert \mathbf{r}, \mathbf{x}, \mathbf{y}, \boldsymbol{\xi}_{-T},\boldsymbol{\Psi})  \propto         \prod_{k=1}^K \left(   n_{s^-,k}^{-T}+\beta  \right)             f\left( r_{ T},x_{ T}, y_T \vert \boldsymbol{\xi}_{T},\boldsymbol{\Psi}  \right).
	\end{equation}
	and 
	\begin{equation}
	f(\boldsymbol{\xi}_{ 0} \vert \mathbf{r}, \mathbf{x}, \mathbf{y}, \boldsymbol{\xi}_{-0},\boldsymbol{\Psi})  \propto \prod_{k=1}^K  \frac{\left(  n_{k,s^+}^{-0}+\beta  \right)  } {\left(  n_{k}^{-0}- \xi_{T,k}+K\beta \right)}. 
	\end{equation}

	It is well known that the MCMC sampler  for HMM tends to mix really slow \citep{Andrieu2010}. To speed up the convergence we try to find an optimal proposal distribution for the Metropolis step, that samples  $K$ $\sigma_1^2$ variables, $K$ $\rho$ variables and $T$ $r$ variables, using the algorithm described in \cite{Robert2009}, page 258. With the goal to speed up the MCMC convergence, 	as a general advice, is suggested to decrease the dimension of the parameters space, i.e. do as much marginalization as possible \citep{Banerjee2003}.  In our model we found convenient  to marginalize over  the vectors $\boldsymbol{\pi}_{k.}, k=1,\dots ,K$ and $\boldsymbol{\pi}_0$ but not over $\mathbf{r}$.  Marginalization over $\mathbf{r}$  decreases significantly the number of random variables to simulate but does not allow to have closed form for full conditional distributions of $\gamma_{k1}$, $\gamma_{k2}$ and $\boldsymbol{\mu}_k$. Without employing the Gibbs step, the MCMC algorithm becomes considerably slower in moving toward its stationary distribution and then the computational burden increases as a larger number of iterations is required. On the other hand, marginalization over $\boldsymbol{\pi}_{k.}, k=1,\dots ,K$ and $\boldsymbol{\pi}_0$ has impact only on the way we simulate $\xi_t, t=0,1,\dots,T$, but their simulation is simple in both cases, with or without $\boldsymbol{\pi}_{k.}, k=1,\dots ,K$ and $\boldsymbol{\pi}_0$, and can be carried out  in a Gibbs step.

	%
	%
	%

	In the estimation step, we  take into account  the label-switching issue, common to all latent-class-based models. This problem occurs when exchangeable priors are used for the state
	specific parameters, which is common practice if there are not prior informations about
	the hidden states. In these cases, the posterior distribution is invariant to permutations
	of the state labels and, hence, the marginal posterior distributions of the state specific
	parameters are identical for all states. Therefore, direct inferences about the state
	specific parameters are not available from the MCMC output.
	Various approaches to deal with the label switching problem in finite mixture models
	have been proposed in the literature; see \cite{jasra2005} for a recent review. To tackle the label switching we decide to use the post processing technique called \emph{pivotal reordering}, proposed in \cite{Spezia2009} or in \cite{marin2013}, Chapter 6.5. 
	
	\subsection{Model Selection}
	
	To decide the number of regimes, we considered the idea of  use the Reversible Jump \citep{GREEN1995} or a non parametric approach, as the one proposed by \cite{Teh04}. However, our main goal is to demonstrate that the CL-GPN is suitable in a HMM Bayesian framework to model circular-linear variables. Thus, we do not want to further increase the complexity of an already highly complex model by introducing $K$ as random variable.
	
	Common model choice criteria are AIC, BIC, ICL and different classification-based
	information criteria which
	are minimized  among a set of potential models. We evaluate these criteria using the  set of parameters, among the MCMC draws, that maximize the posterior distribution  (called  maximum at posterior, MAP  or  MAP estimator) \cite[][Section 4.4.2, 7.1.4]{fruhwirthschnatter2006}. 
	Let $\tilde{\boldsymbol{\Psi}}$ be the MAP estimator, we compute the $BIC$ and $AIC$ as  
	$
	BIC =-2 \log \left( f\left(  \mathbf{x},	\mathbf{y} |\tilde{\boldsymbol{\Psi}}    \right)  \right)+ \# parameters\times \log(T)
	$
	and   
	$
	AIC =-2  \log \left(f\left(  \mathbf{x},	\mathbf{y} |\tilde{\boldsymbol{\Psi}}   \right)   \right)+ 2\times \# parameters.
	$

	The BIC and AIC are generally criticized since they do not take into account the quality of  classification of the variables in the $K$ regimes. For classification purpose \cite{Biernacki:2000} propose to use the ICL; an index based on the likelihood of observed variables and the vector of regimes indicator that is used by \cite{Celeux:2008} in a HMM context. We compute a BIC approximation of the $
	ICL =  f(\mathbf{x},\mathbf{y}|\tilde{\boldsymbol{\xi}} ,\tilde{\boldsymbol{\Psi}})- 2 \log  f (\tilde{ \boldsymbol{\xi}})  + \# parameters\times \log(T), 
	$ \cite[see for example ][pag. 214]{fruhwirthschnatter2006} 
	
	in the latter case, as suggested by \cite{MC2000}, pag. 216,  we first obtain an estimator of $ \boldsymbol{\xi}$, i.e. the MAP $\tilde{\boldsymbol{\xi}}$, and then, as for the BIC and AIC,  we compute the ICL using  the MAP  estimator of ${\boldsymbol{\Psi}}$  conditioning on  the value $\tilde{\boldsymbol{\xi}}$.


	\section{SIMULATION STUDY}
	\label{s:ss}

	In this Section we carried out a simulation study to investigate the performance of the proposed  approach
	in recovering model parameters and the hidden structure of the data. We empirically demonstrate that the CL-GPN can be used in presence of both unimodal or bimodal state-dependent circular distributions and that ignoring the dependencies between the circular and linear variable at a given time leads to a higher number of states.

	We plan the simulation study to cover schemes with different underlying {\it null} models assuming bimodal or almost uniform shapes for the circular variable,  and overlapping or well-separated state-dependent distributions for the linear variable. 
	On each simulated datasets, we estimate three models: 1) the CL-GPN model; 2)  a constrained model, defined as diagonal CL-GPN (CL-DPN), with $\boldsymbol{\Sigma}_k=\mathbf{I}_2$, so that the state-dependent circular distribution is symmetric and  unimodal; 3) a  CL-GPN model with all the $\gamma_{k1}$ and $\gamma_{k2}$ equal to zero, i.e. assuming independence between circular and linear variable given the latent state (indicated as Ind-CL-GPN). 
	
	%

	\subsection{Designing the simulation study}
	
	For each null model, we simulated 200 datasets considering two time-series lengths, $T = 500$ and $ T=2000$, with $K=3$,  $\boldsymbol{\xi}_0=1$ and  transition matrix $\boldsymbol{\pi}$ with  diagonal elements equal to 0.8 and extradiagonal elements equal to 0.1. 
	The considered schemes are summarized in Figure \ref{fig:dist}, and are characterized by the following settings:
	\begin{figure}[t!]
		\centering
		\includegraphics[scale=0.5]{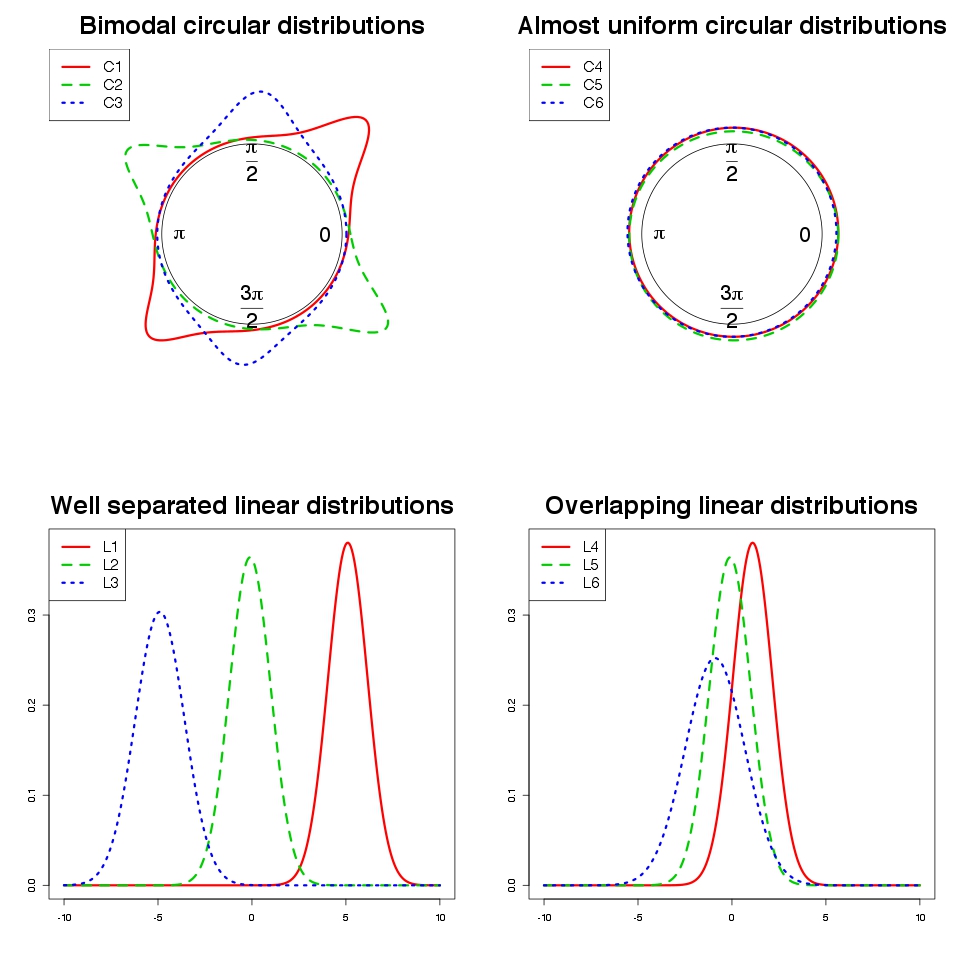}
		\caption{Marginal distributions used in the simulation examples.}\label{fig:dist}
	\end{figure}

	\begin{figure}[t!]
		\centering
		\includegraphics[scale=0.5]{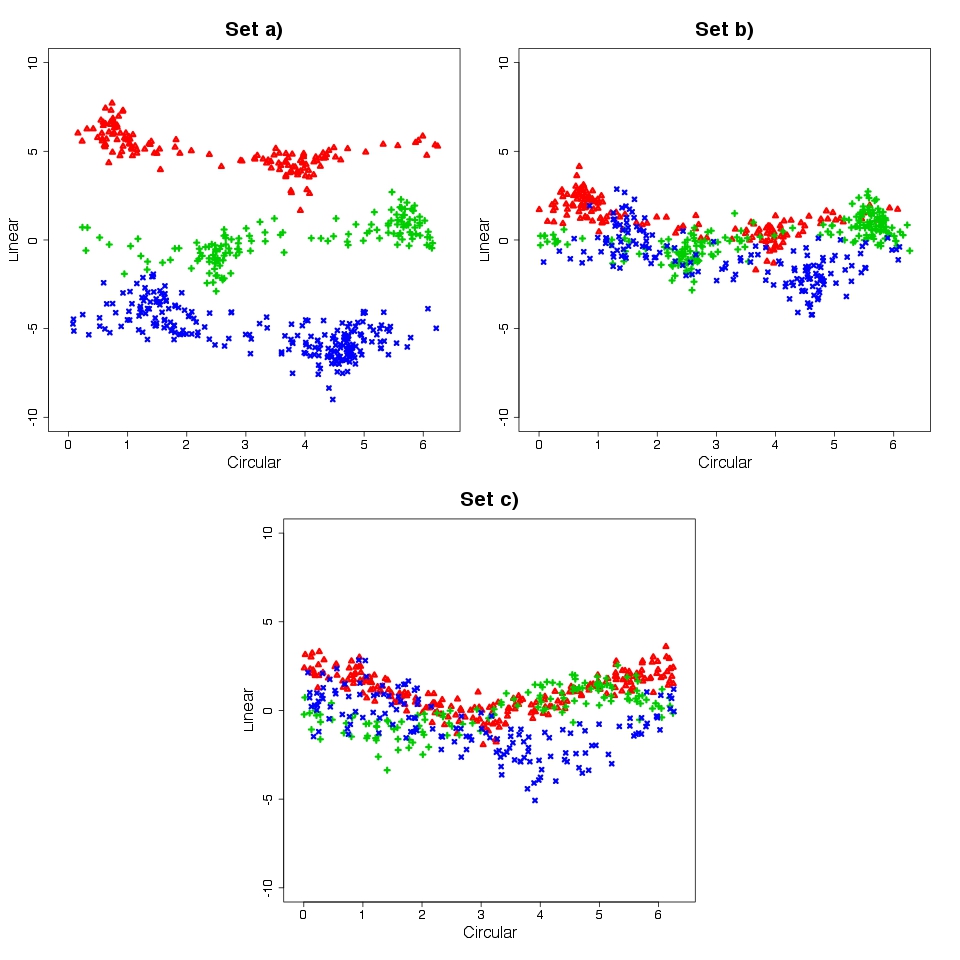}
		\caption{Scatter plot of one simulated dataset for each set of parameters $(T=500)$.  }\label{fig:scatter}
	\end{figure}
	\begin{enumerate}
		\item[a)] distributions C1 and L1,  C2 and L2 and C3 and L3 are considered as state-dependent distributions for the first, the  second and the third regime, respectively. The joint representation through scatters is displayed in Figure \ref{fig:scatter}. This scheme has bimodal state-dependent circular distributions and well separated linear ones. The following parameters are	used to generate data:
		\begin{equation*}
		\boldsymbol{\mu}=\left[ \begin{array}{lll}
		\mu_{11}& \mu_{12} & \mu_{13}\\
		\mu_{21} & \mu_{22} & \mu_{23}
		\end{array} \right]=
		\left[ \begin{array}{ccc}
		0.1 & 0.1 & 0.0\\
		0.1 & -1.0 &-0.1
		\end{array} \right]
		\end{equation*}\quad\begin{equation*}
		\boldsymbol{\gamma}=\left[ \begin{array}{lll}
		\gamma_{01}& \gamma_{02}& \gamma_{03}\\
		\gamma_{11} & \gamma_{12}& \gamma_{13}\\
		\gamma_{21} & \gamma_{22}& \gamma_{23}\\
		\end{array} \right]=
		\left[ \begin{array}{ccc}
		5.0 & 0.0& -5.0\\
		1.0 & 0.0& 1.0\\
		0.0 & -1.0&1.0
		\end{array} \right]
		\end{equation*}
		$$ \sigma^{2}_{k1}=\left\{\begin{array}{cc}1 & k =1\\ 2&k=2\\0.1 & k =3 \end{array}\right.; \quad\sigma^{2}_{ky}=\left\{\begin{array}{cc}0.1 & k =1\\ 0.2 & k=2\\0.5 & k =3 \end{array}\right.;\quad\rho_{k}=\left\{\begin{array}{cc}0.9 & k =1\\-0.9&k=2\\ 0.2 & k =3 \end{array}\right.$$
		\item[b)] this setting shares circular distributions with scheme a) while the state-dependent linear distributions are respectively the density L4, L5 and L6 of Figure \ref{fig:dist}. The joint representation through scatters is displayed in Figure \ref{fig:scatter}. With respect to scheme a), we change the  values of $\boldsymbol{\gamma}$:
		\begin{equation*}
		\boldsymbol{\gamma}=\left[ \begin{array}{lll}
		\gamma_{01}& \gamma_{02}& \gamma_{03}\\
		\gamma_{11} & \gamma_{12}& \gamma_{13}\\
		\gamma_{21} & \gamma_{22}& \gamma_{23}\\
		\end{array} \right]=
		\left[ \begin{array}{ccc}
		1.0 & 0.0& -1.0\\
		1.0 & 0.0& 1.0\\
		0.0 & -1.0&1.0
		\end{array} \right]
		\end{equation*}
		to have more overlapping state-dependent distributions for the linear variable. 
		\item[c)] the state-dependent distributions for the linear variable are the same as in scheme b), whilst the circular ones are respectively the density C4, C5 and C6 of Figure \ref{fig:dist}. The joint representation through scatters is displayed in Figure \ref{fig:scatter}.
		In this case we simulate from a CL-DPN since we use  $\sigma^2_{11} = \sigma^2_{12}=\sigma^2_{13}=1$ and $\rho_1=\rho_2=\rho_3 = 0$, i.e. the circular variable  has state-dependent unimodal (almost uniforms) distributions. 
	\end{enumerate}
	
	On each dataset we estimate models with $K$ from 2 to 6 and assuming the following prior distributions: $\mu_{ki} \sim N(0,5)$, $\gamma_{kj} \sim N(0,5)$, $\rho_k \sim N(0,5)I(-1,1)$, $\sigma_{k1}^2 \sim IG(2,1) \footnote{The two parameters are the shape and rate, respectively}$,  $\sigma_{ky}^2 \sim IG(2,1)$, $\beta=1$ with  $i=1,2 $ and $j=1,2,3$; i.e. they do not depend on the regime.
	
	\subsection{Simulation Study Results}
	To evaluate the performance of AIC, BIC and ICL as selection criteria for the number of regimes, in Table \ref{tab:aicbicicl} we report the frequency distribution of the predicted $K$ under each simulation setting considered for the CL-GPN, CL-DPN and Ind-CL-GPN models. With respect to the CL-GPN model, we can observe that ICL performs considerably well in all cases. In fact, the predicted $K$ is only occasionally different from the true one, and, when this happens, the former is always larger than the latter. On the other hand, AIC and BIC have an excellent behavior with the exception of the cases $T=500$, schemes a) and b). As may be expected, these criteria perform better as the amount of information in the data increases.
	
	Ignoring the (state-dependent) correlation between circular and linear measurements may strongly affect the hidden structure. Indeed, by looking at information criteria for the Ind-CL-GPN model, we have that the latent structure is not well recovered and a higher number of regimes than expected is estimated. Of course, this affects parameter estimates and results interpretation, as a not needed number of (latent) regimes is identified in the data.

	\begin{table}[t!]
		\centering
		\tiny{
			\caption{Frequency distribution of predicted number of regimes } \label{tab:aicbicicl}
			\begin{tabular}{ccc|ccccc||ccccc||ccccc}
				& & & \multicolumn{5}{c||}{Predicted $K$ (AIC)}& \multicolumn{5}{c||}{Predicted $K$ (BIC)}& \multicolumn{5}{c}{Predicted $K$ (ICL)}\\
				T & Model & Scheme & 2 & 3 & 4 & 5 & 6& 2 & 3 & 4 & 5 & 6& 2 & 3 & 4 & 5 & 6 \\ \hline
				500  & CL-GPN & a)   &  0.00 &  0.62 &  0.35 &  0.03 &  0.00  &  0.00 &  0.82 &  0.18 &  0.00&  0.00    &  0.00 &  0.91 &  0.09 &  0.01&  0.00   \\
				500  & CL-GPN & b)   &  0.00 &  0.68 &  0.25 &  0.07&  0.00  & 0.00 &  0.67 &  0.30 &  0.03&  0.00    &  0.00 &  0.90 &  0.09 &  0.01&  0.00   \\
				500  & CL-GPN & c)   &  0.00 &  0.98 &  0.02 &  0.00&  0.00  &  0.00 &  0.98 &  0.02 &  0.00&  0.00    &  0.00 &  0.99 &  0.01 &  0.00&  0.00   \\ \hline
				500  & CL-DPN & a)   &  0.00 &  0.59 &  0.37 &  0.04 &  0.00  &  0.00 &  0.80 &  0.18 &  0.02&  0.00    &  0.00 &  0.87 &  0.10 &  0.03&  0.00   \\
				500  & CL-DPN & b)   &  0.00 &  0.61 &  0.31 &  0.08&  0.00  & 0.00 &  0.63 &  0.33 &  0.04&  0.00    &  0.00 &  0.86 &  0.12 &  0.02&  0.00   \\
				500  & CL-DPN & c)   &  0.00 &  0.98 &  0.01 &  0.01&  0.00  &  0.00 &  1.00 &  0.00 &  0.00& 0.00   &  0.00 &  1.00 &  0.00 &  0.00&  0.00   \\ \hline	
				500  & Ind-CL-GPN & a)   & 0.00 & 0.00 &  0.01 & 0.08 & 0.91  &  0.00 & 0.00 &  0.00 &   0.09& 0.91    &  0.00 &  0.00 &  0.08 & 0.26& 0.66  \\
				500  & Ind-CL-GPN & b)  &  0.00 &  0.00 &  0.02  & 0.08 & 0.90   &  0.00 &  0.00 &  0.05 & 0.03 & 0.92    &   0.03 &0.00 &0.00 & 0.07& 0.90   \\
				500  & Ind-CL-GPN & c) &   0.00 & 0.00 & 0.00 &  0.09& 0.91  &   0.00 & 0.00 &  0.00 &  0.09 &0.91    &  0.48& 0.41& 0.06 &  0.03&  0.02     \\ \hline			
				2000  & CL-GPN & a)   &  0.00 & 0.97 &  0.03 &  0.00&  0.00  &  0.00 &  0.98 &  0.01 &  0.01&  0.00    &  0.00 &  0.98 &  0.00 &  0.00&  0.00   \\
				2000  & CL-GPN & b)   &  0.00 &  0.97 &  0.01 &  0.02&  0.00  &  0.00 &  0.98 &  0.02 & 0.00&  0.00    &  0.00 &  0.98 & 0.02 &  0.00&  0.00   \\
				2000  & CL-GPN & c)   & 0.00 &  0.96 &  0.03 &  0.01&  0.00  &  0.00 &  0.99 &  0.01 &  0.00&  0.00    & 0.00&  1.00 &  0.00 &  0.00&  0.00   \\ \hline
				2000  & CL-DPN & a)   &  0.00 & 0.90 &  0.08 &  0.02&  0.00  &  0.00 &  0.91 &  0.05 &  0.04&  0.00    &  0.00 &  0.93 &  0.07 &  0.00&  0.00   \\
				2000  & CL-DPN & b)   &  0.00 &  0.91 &  0.07 &  0.02&  0.00  &  0.00 &  0.93 &  0.03 & 0.04&  0.00    &  0.00 &  0.91 & 0.09 &  0.00&  0.00   \\
				2000  & CL-DPN & c)   &  0.00 & 0.98 &  0.02 &  0.00&  0.00  &  0.00 &  1.00 &  0.00 & 0.00&  0.00    &  0.00 &  1.00 &  0.00 &  0.00&  0.00   \\ \hline	
				2000  & ind-CL-GPN & a)   &  0.00 &  0.00 &  0.02 & 0.31 & 0.67  &  0.00 &  0.00 &  0.02 & 0.27 & 0.71   &  0.00 &  0.04 &0.21 &0.31& 0.44   \\
				2000  & ind-CL-GPN & b)   & 0.00 &  0.00 &   0.03&  0.07 &0.90   &  0.00 &  0.00 &  0.02 & 0.08& 0.90      &  0.03 & 0.06& 0.22& 0.41& 0.28   \\
				2000  & ind-CL-GPN & c)&0.00    &  0.00 &  0.00 &  0.29 &0.71&  0.00  &  0.00 & 0.00 &  0.25 &0.75    &  0.06& 0.27 &0.39 &0.24&  0.04  \\ \hline
			\end{tabular}
		}
	\end{table}
	
	Here we briefly summarize the results of the simulation study for scheme c). 
	By looking at parameters estimates (see Table \ref{tab:scheme9}), we have that the CL-GPN and the CL-DPN models lead  essentially to  the same results. Point estimates and credibility intervals  are very close, suggesting that the CL-GPN distribution can be used whenever we cast doubts on the unimodality of circular distributions. Indeed, the CL-DPN distribution is a specific case of the CL-GPN one, in which conditional circular distribution are constrained to be unimodal. 
	
	\begin{table}[t!]
		\centering
		\small{
			\caption{posterior median estimates of the parameter ( $\hat{}$ ) and credibility intervals (CI)  for scheme c) and $T=500$} \label{tab:scheme9}
			\begin{tabular}{lccc|ccc}
				\hline \hline
				&  & CL-GPN &   & & CL-DPN & \\ \hline 
				& $k=1$ & $k=2$ & $k=3$  & $k=1$ & $k=2$ & $k=3$ \\ \hline
				$\hat{\mu}_{k1}$ &  0.12 & 0.12 & -0.05 &  0.13 & 0.11& -0.06 \\
				CI		&  (-0.04  0.28)    & (-0.09  0.33)  &(-0.26  0.15) &  (-0.03  0.31)    & (-0.11  0.34)  &(-0.27  0.16)\\
				$\hat{\mu}_{k2}$ &  0.04 &  -0.08   & 0.07 &0.03   &  -0.09   & 0.09 \\
				CI			& (-0.12  0.19)  & (-0.30  0.12)  & (-0.14  0.27) & (-0.13  0.18)  & (-0.32  0.14)  & (-0.12  0.29)\\
				$\hat{\rho}_k$ &   0.06  &   -0.09  & 0.14 &  $\cdot$ &  $\cdot$   & $\cdot$ \\
				CI  & ( -0.13  0.22 )   & (-0.32  0.16)&  (-0.11  0.37) & ($\cdot$ $\cdot$)  & ($\cdot$ $\cdot$)  & ($\cdot$ $\cdot$)\\
				$\hat{\sigma}_{k1}^2$ & 0.94   &  0.76   & 1.01 &  $\cdot$ &  $\cdot$   & $\cdot$ \\
				CI & (0.65  1.35)  & (0.47  1.22)  & (0.63  1.57)  & ($\cdot$ $\cdot$)  & ($\cdot$ $\cdot$)  & ($\cdot$ $\cdot$)\\
				$\hat{\gamma}_{k0}$ & 0.98    &  -0.04   & -1.04  & 0.98   &   -0.04  &   -1.05 \\
				CI & (0.89  1.06)   & (-0.19  0.13) & (-1.22 -0.86)  & (0.89  1.06)  & (-0.20  0.12)  & (-1.23 -0.88)\\
				$\hat{\gamma}_{k1}$ &  1.04   & 0.14  & 0.82 &  1.01 &   0.15  & 0.86 \\
				CI  &  (0.88  1.22)  & (-0.04  0.35) &  (0.60  1.07)  & (0.91  1.13)  & (-0.01  0.33)  & (0.68  1.06)\\
				$\hat{\gamma}_{k2}$  & -0.04  & -1   & 1.04  & -0.02  &   -0.96  & 1.05 \\
				CI &  (-0.11  0.05)  & (-1.18 -0.83)  & (0.82  1.28)  & (-0.09  0.05)  & (-1.12 -0.81)  & (0.87  1.26) \\
				$\hat{\sigma}_{ky}^2$  &  0.14   &  0.32   & 0.42 & 0.14 & 0.3    &  0.41 \\
				CI & (0.09  0.19) & (0.18  0.52) & (0.26  0.67)  & (0.10  0.19)  & (0.18  0.49)  & (0.25  0.63) \\ \hline 
			\end{tabular}
		}
	\end{table}
	
	To further resemble empirical situations, we randomly drop 10\% observation of a randomly selected dataset simulated accordingly to the scheme c) with $T=500$ and estimate the CL-GPN and a CL-DPN models. Along with model parameters, we also simulate  the missing observations. We compute the average continuous ranked probability score (CRPS) for both the circular \citep{grimit2006} and linear variable \citep{gneiting22} from the posterior samples of the missing observations, as well as the average prediction error (APE) for the circular variable \citep{Jona2013} and the mean squared error for the linear ones (MSE). With the CRPS we evaluate the model performance regarding the entire predictive distribution. APE and MSE allow us to measure the distance between the true values and the simulated ones. The CPRPs for the circular variable and the MSE are identical under the two models while the CRPS for the linear one is 0.66 under the CL-GPN and 0.67 under the CL-DPN  and the APE is respectively 0.76 and 0.75 for the CL-GPN and the CL-DPN. Then the two models have the same performances in dealing with the missing values as well. 
	
	From the computational point of view, in the datasets with $T=2000$ our C++ implementation of the model needs $1000000$ iterations with a burnin of $700000$ and a thin of $100$ while with $T=500$ the iterations needed are $800000$, with a burnin of $400000$ and again a thin  of  $100$.
	The computational work for the simulation study and the real data application of Section  \ref{s:real}, has been executed on the IT resources made available by ReCaS, a project financed by the MIUR (Italian Ministry for Education, University and Re-search) in the ``PON Ricerca e Competitività 2007-2013 - Azione I - Interventi di rafforzamento struttural'' PONa3\_00052, Avviso 254/Ric.  The computational time are of the order of 1 hour for $T=500$ and $5$ hours for $T=2000$.
	All the results shown are from MCMC chains that reach the convergence, checked using the standard tool on the R package coda.\\

	\section{REAL DATA EXAMPLE}
	\label{s:real}
	
\begin{figure}[t!]
		\centering
		\includegraphics[scale=0.5]{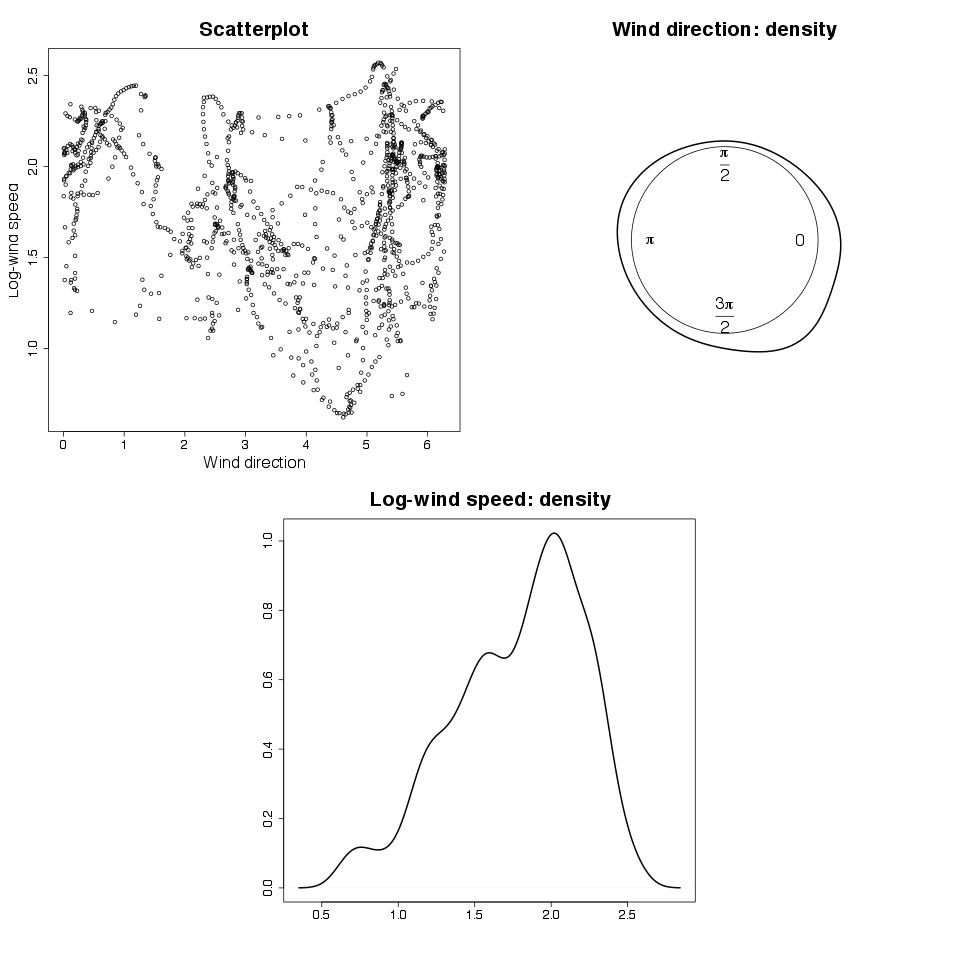}
		\caption{Real data. } \label{fig:realplot}
	\end{figure}

	Finally, we apply the CL-GPN hidden Markov model to a bivariate time series of wind directions and (log-transformed) speeds. Data are recorded on a semi-hourly base from 12/12/2009 to 12/1/2010 in Ancona (Italy) at a bouy located in the Adriatic Sea 30km from the coast (see Figure \ref{fig:realplot}). Data are recorded on $T=1500$ times and have been previously analysed by \cite{Bulla2012}.

	As often arise in environmental studies, data are not complete.  213 and 210 missing values are recorded for directions and speeds, respectively; 125 profiles are completely missing.
	
	During wintertime, relevant wind
	events in the Adriatic Sea are typically generated by the south-eastern Sirocco, the north-eastern
	Bora and the north-western Maestral. Sirocco arises
	from a warm, dry, tropical air mass that is pulled northwards by low-pressure cells moving
	eastwards across the Mediterranean Sea. By contrast, Bora episodes occur when a polar
	high-pressure area sits over the snow-covered mountains of the interior plateau behind the
	coastal mountain range and a calm low-pressure area lies further south over the warmer
	Adriatic. Finally, the Maestral is a sea-breeze wind blowing northwesterly when the east
	Adriatic coast gets warmer than the sea. While Bora and Sirocco episodes are usually
	associated with high-speed flows, Maestral is in general linked with good meteorological
	conditions. Hence, the marginal distribution of (log-transformed) wind speed may
	be interpreted as the result of mixing different wind-speed regimes.

\begin{figure}[t!]
		\centering
		\includegraphics[scale=0.5]{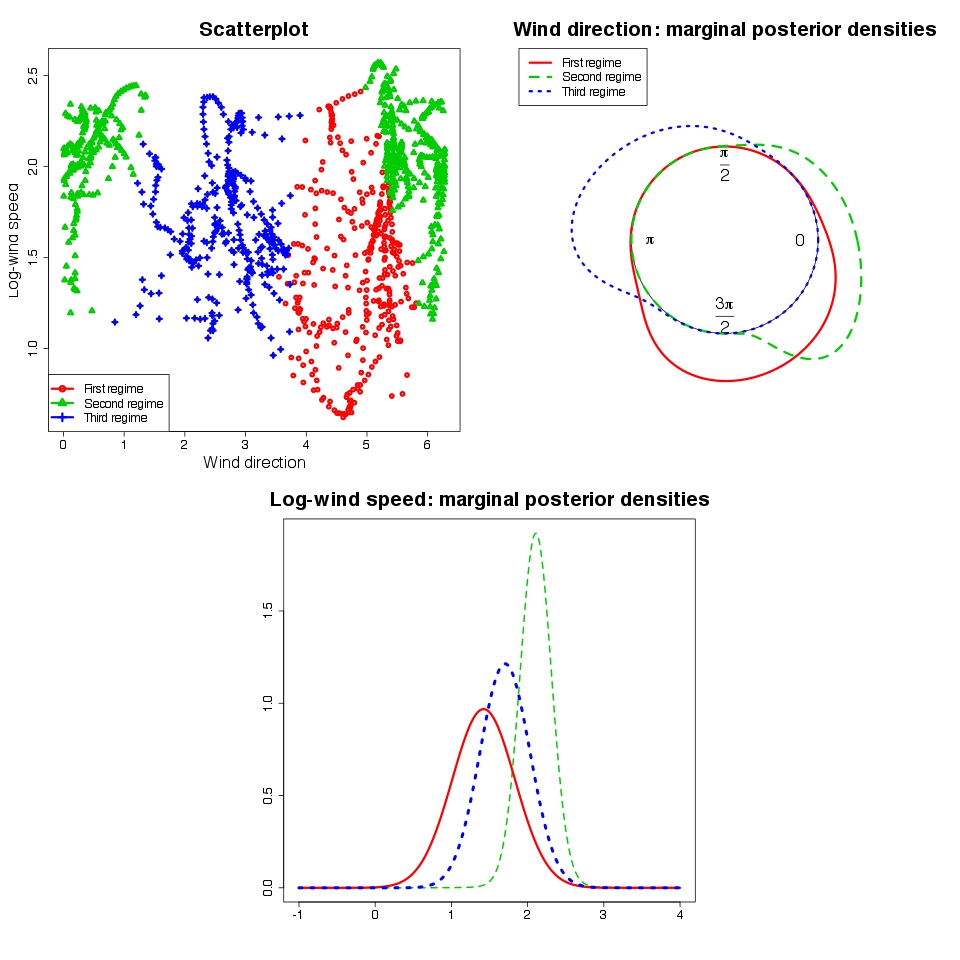} 
		\caption{Real data classification. } \label{fig:density}
	\end{figure}

	As for the simulation examples, we look at the AIC, BIC and ICL to select the appropriated number of components. The ICL suggest to use $K=3$ while the AIC and BIC $K=4$. To help decide between the two number of regimes, we look ar their  predictive ability,  the CRPS$_c$ and APE highlight loss  of predictive ability on the circular variable if we choose $K=4$ (CRPS$_c$=0.59 and APE=0.75 with $K=4$ while CRPS$_c$=0.34 and APE= 0.35 with $K=3$). For the linear variable, looking at the values of CRPS$_l$ and MSE, there is a small difference between  $K=3$ and $K=4$ however both  CRPS$_l$ and MSE favour  $K=3$ (CRPS$_l$=0.17 and APE=0.39 with $K=4$ while CRPS$_l$=0.16 and APE= 0.34 with $K=3$). We  decide to adopt $K=3$, that is also the choice of  \cite{Bulla2012} following their suggestion that three regimes provide well-separated and more interpretable states. The resulting classification is displayed in Figure \ref{fig:density} and  all the credibility intervals and point estimates of the parameters are in Table \ref{tab:realste}. The estimated transition probabilities are displayed in Table \ref{tab:realtransmat}. As expected, the transition probability matrix
	is essentially diagonal, reflecting the temporal persistence of the regimes, i.e. of wind conditions . Furthermore,
	the small off-diagonal transition probabilities between states indicate that direct transitions between Sirocco and Bora episodes are very unlikely. The model hence confirms that the Adriatic Sea typically alternates relevant wind events with periods of good conditions.
	
	\begin{table}
		\centering
		\small{
			\caption{Real data: posterior median estimates of the parameter ( $\hat{}$ ) and credibility intervals (CI)} \label{tab:realste}
			\begin{tabular}{cccc}
				\hline \hline
				& $k=1$ & $k=2$ & $k=3$ \\ \hline
				$\hat{\mu}_{k1}$ & 0.45  & -1.62& 1.19 \\
				CI		&  (0.28  0.63)    & (-1.83 -1.41 )  &(1.02 1.32 ) \\
				$\hat{\mu}_{k2}$ & -1.80  &   0.67  &  -0.41\\
				CI		& (-2.07  -1.57)  & (0.52  0.81 )  & (-0.51 -0.30) \\
				$\hat{\rho}_k^2$ &   0.36  &   0.56  &  -0.27\\
				CI & ( 0.14  0.54 )   & (0.38  0.71 )&  (-0.46 -0.06 ) \\
				$\hat{\sigma}_{k1}^2$ & 2.03   &   0.94   &  0.15\\
				CI & (1.46  2.85 )  & (0.71  1.28 )  & (0.10   0.23)\\
				$\hat{\gamma}_{k0}$ &   1.34  &   1.47  &   2.36\\
				CI & (1.21  1.49)   & (1.33  1.62 ) & ( 2.25  2.45)\\
				$\hat{\gamma}_{k1}$ &   0.01   &  -0.09 &  -0.22\\
				CI  &  (-0.03  0.05 )  & (-0.16 -0.03) &  (-0.30 -0.13)\\
				$\hat{\gamma}_{k2}$  &  -0.04 &   0.12 &  -0.02 \\
				CI &  (-0.11  0.03 )  & (0.06  0.18 )  & (-0.05  0.01)  \\
				$\hat{\sigma}_{ky}^2$  &   0.17  &   0.09  &   0.03\\
				CI & ( 0.14  0.19) & (0.08  0.11 ) & (0.03  0.06)  \\ \hline   
			\end{tabular}
		}
	\end{table}
	
	\begin{table}
		\centering
		\small{
			\caption{Real data: Transition matrix } \label{tab:realtransmat}
			\begin{tabular}{lcccc}
				\hline \hline
				Destination state &  & 1 & 2 & 3 \\ \hline
				&  1 &   0.96 &    0.02 &   0.02\\
				&& (0.94,  0.97)  & (0.01,  0.04)  & ( 0.01,  0.04)\\
				Origin state   
				&   2 &   0.02 &   0.97 &   0.00 \\
				&&  (0.01  0.04)   & ( 0.95,  0.98)  & (0.00,  0.01)\\
				&   3 &   0.02 &   0.00 &   0.98 \\
				&& (0.01,  0.03) & (0.00,  0.01) & (0.97,  0.99) \\     \hline 
			\end{tabular}
		}
	\end{table}

	For a more clear interpretation of the state dependent distributions, we compute some feature of the CL-GPN distribution. In detail, we look at the posterior marginal mean and variance of the linear distribution  for each regime ($\bar{\mu}_{ky} $ and  $\bar{\sigma}_{ky}^2$), the circular mean ($\bar{\mu}_{kx}$) and concentration
	($\bar{g}_{kx}$) of the circular variable and a measure of correlation between the circular and linear variables ($\bar{\rho}_{kxy}^2$) as in \cite{Merdia1999}, pag.  245. Point estimates and credibility intervals are provided in Table \ref{tab:realpost}).

	\begin{table}
		\centering
		\small{
			\caption{Real data:  posterior median estimates  ( $\hat{}$ ) and credibility intervals (CI)     of the features of the distribution CL-GPN} \label{tab:realpost}
			\begin{tabular}{cccc}
				\hline \hline
				& $k=1$ & $k=2$ & $k=3$ \\ \hline
				$\hat{\bar{\mu}}_{kx}$ & 4.98  & 2.7 &  6.04\\
				CI		&  (4.81  5.16)    & (2.55  2.83)  &(5.90  6.19) \\
				$\hat{\bar{g}}_{kx}$ & 0.78  & 0.83 & 0.82 \\
				CI		&  (0.71  0.84)    & (0.77  0.87)  &(0.76  0.86) \\
				$\hat{\bar{\mu}}_{ky}$ &1.42   & 1.70 & 2.11 \\
				CI			&  (1.34  1.50)    & (1.63  1.78)  &(2.02  2.20) \\
				$\hat{\bar{\sigma}}^2_{ky}$ & 0.17  & 0.11 & 0.04 \\
				CI		&  (0.15  0.20)    & (0.09  0.12)  &(0.03  0.07) \\
				$\hat{\bar{\rho}}_{kxy}^2$ &0.02   &0.05 &  0.05\\
				CI		&  (0.00   0.10)    & (0.00  0.18)  &(0.00  0.17) \\ 			 \hline 
			\end{tabular}
		}
	\end{table}

	The regimes are ordered accoridng to the marginal log-wind speed. In the three regimes the point estimates are respectively $\hat{\mu}_{ky} = 1.42, 1.70, 2.11$  (which correspond to $4.14m/s,$ $5.47m/s,8.25m/s$ in the natural scale). 
	With the increases of the velocity, the distribution becomes more concentrated: the marginal linear variance,  $\hat{\bar{\sigma}}_{ky}^2$, is respectively $0.17, 0.11,0.04$, for a plot of the distributions see Figure \ref{fig:density}. The circular mean is 4.98 in the first regime, north-westerly
	Maestral episodes,  2.70 in the second, south-eastern Sirocco, and 6.04 in the third, northern Bora jets,  on the first regime the circular marginal distribution is less concentrated than in the others (0.78 for $k=1$,  0.83 for $k=2$ and 0.82 for $k=3$). 
	
	The correlations between the circular and linear variables are weak in all the regimes: $\hat{\bar{\rho}}^2_{kxy}$ is 0.02 in the first and 0.05 in the others.
	Under the hypothesis of no correlation, i.e. ${\bar{\rho}}^2_{kxy}=0$, the statistic $\tilde{F}=\frac{{\bar{\rho}}^2_{kxy}(n-1)}{1-{\bar{\rho}}^2_{kxy}}$ is distributed as a $F_{2,T-3}$ where in our case $T=1500$. The lower limits of the $95\%$ credibility intervals of the posterior distributions  of $\tilde{F}$ are 1.24, 4.80 and 4.95  in the calm, transition and storm conditions respectively, and the $95 \%$ percentile of $F_{2,T-3}$ is 3.00. Accordingly, circular-linear correlations are significant in the transition and storm conditions only. This result is not present at all in previous analyses. This can be seen also with the value of $\gamma_{k1}$ and $\gamma_{k2}$ in Table \ref{tab:realste}. In the first regime $\hat{\gamma}_{11} = 0.01$ and  $\hat{\gamma}_{12} = -0.04$, both credibility intervals contain the 0. In the second there is a negative relation between the linear variable and the cosine of the circular one ($\gamma_1=-0.09 $) and a positive relation with the sine ($\gamma_2 = 0.12$). In the third regime the dependence between the linear and circular variable is on the cosine direction ($\gamma_1 = -0.22$).

	We estimate the model using  $1000000$ iterations, a burnin of $700000$ and a thin of 100. Here again we checked the convergence of the MCMC chain using the standard tool on the R package coda.
	
	\section{Discussion}
	\label{s:discussion}
	
	In this work we introduce, for the first time, the CL-GPN distribution in a Bayesian HMM framework and we present the explicit expression of the CL-GPN likelihood. This approach allows to easily model multivariate processes with mixed support (circular-linear), by combining the bivariate representation of the circular component (i.e. the projected normal distribution) and a Gaussian distribution for the linear part. Here we considered one circular and one linear variable although it is  fairly easy to extend the proposed model to more than one linear component.
	
	The Bayesian framework allows us to overcome identifiability issues and computational problems that may arise in the classical setting. Several implementation novelties are introduced to speed up algorithms convergence. We use an adaptive Metropolis whenever a Gibbs sampler is not implementable (section \ref{sec:comp}). Furthermore we marginalize the transition matrix so to avoid its estimation to reduce the problem size obtaining it as an a posteriori byproduct (Section \ref{sec:postinf}) and we provide evidence that the marginalization does not affect parameters estimation.  We also demonstrate that assuming conditional independence between the circular and linear variable can make difficult to correctly estimate the number of regimes. 
	
	We applied this methodology to wind data confirming previously obtained results and highlighting new data features.
	Circular parameters interpretation is not straightforward, however this does not limit the inferential richness of the model. Using MCMC simulations posterior circular mean and concentration can be derived, as well as the circular-linear correlation. Of course, different areas of application can be considered for the proposed approach, e.g. animal movement modelling \citep{Langrock2014} and driving behaviour \citep{jackson2014}.

	Further developments will include the extension to more than one circular variable. This extension requires a careful definition of correlation between circular variables that is not straightforward under the projected normal distribution. Another interesting extension of the proposed approach is to allow the estimation of the number of states along with the model parameters. The latter can be obtained using a hierarchical Dirichlet process on the states or a reversible jump.

	A crucial assumption of our model is that the temporal dependence is well described by a first order Markov chain, i.e. the sojourn time is geometrical. If we want to allow for different sojourn time distributions with finite support the HMM formulation is exact. Similarly by allowing the number of hidden states to grow with the sample size, we can allow for continuos time, i.e. the hidden distribution can be approximated with arbitrary accuracy using the proposed model.  This can be seen as a possible solution to computational issues arising with continuous-valued latent models \citep{Langrock2012a}.

	
	%

	\bibliographystyle{wb_env} 
	\bibliography{allHMM}
	\end{document}